# Mild-to-wild plasticity of Earth's upper mantle

David Wallis[1], Kathryn M. Kumamoto[2], Thomas Breithaupt[1]

[1]Department of Earth Sciences, University of Cambridge, Downing Street, Cambridge, U.K., CB2 3EQ.

[2]Lawrence Livermore National Laboratory, Livermore, California, U.S.A.

**The flow of Earth's upper mantle has long been considered to occur by slow and near-continuous creep. Such behaviour is observed in classical high-temperature deformation experiments[1,2] and is a fundamental component of geodynamic models[3]. However, the latest generation of high-resolution experiments, capable of sensing temporal heterogeneities in rates of dislocation motion, have revealed that materials ranging from metals to ice exhibit a spectrum of behaviours termed mild-to-wild plasticity[4–6]. It remains unknown whether olivine, the most abundant mineral in Earth's upper mantle, always exhibits mild continuous flow or can exhibit intermittent wild fluctuations in plastic strain rate. Here, we demonstrate that olivine exhibits measurable wildness, even under conditions at which its behaviour is predicted to be relatively mild. During nanoindentation experiments conducted at room temperature, continuous plastic flow is punctuated by intermittent bursts of displacement with log-normally distributed magnitudes, indicating avalanches of correlated dislocation motion that account for approximately 8 ± 6% of the plastic strain. Remarkably, the framework of mild-to-wild plasticity predicts that wildness should increase with depth in the Earth, with flow of the asthenospheric upper mantle occurring almost entirely by wild fluctuations of deformation at the grain scale. The recognition of intermittent plasticity in geological materials provides new constraints for microphysical models of dislocation-mediated flow and raises questions about the mechanisms of transient instabilities in otherwise ductile regimes, such as deep earthquakes and slow-slip events.**

The flow of hot but solid rocks in Earth's mantle at the base of the tectonic plates is an archetypal slow and steady process. The motion of Earth's surface appears smooth and continuous when observed by geodetic techniques[7], implying that the associated flow of hot, ductile rocks at the base and margins of each plate is similarly steady, at least when observed at large spatiotemporal scales. Intermittent bursts of deformation do occur in the form of earthquakes, mostly in the cold, brittle veneer of Earth's upper crust, yet the viscous deformation of deeper, hotter rocks that they induce in their aftermath also appears as smooth flow when detected at the surface[8,9]. These observations are mirrored in classical plasticity and creep experiments, in which millimetre- or centimetre-sized crystals flow steadily[10–12]. However, it cannot be taken for granted that these observations accurately characterise the nature of flow at and below the grain scale in the upper mantle, nor that they provide a complete picture of the characteristics of flow at larger scales. For instance, some sets of geophysical observations have been postulated to result from transient instabilities of deformation under conditions typically associated with viscous flow. For example, some earthquakes in subduction zones occur at depths where the confining pressure is expected to suppress frictional failure and instead have long been hypothesised to result from thermal runaway during viscous flow[13,14]. More recently, slow-slip events, which are enigmatic phenomena involving deformation at rates intermediate between those of background levels of creep and seismogenic earthquakes, have also been suggested to occur due to transient instabilities of viscous flow resulting from accelerated dislocation glide[15]. These observations motivate a reexamination of the fundamental nature of viscoplastic flow in geological materials.

Recent developments in the material sciences have led to recognition of a spectrum of deformation behaviours termed mild-to-wild plasticity[5,6,16]. Advances in analytical techniques with high spatial and/or temporal resolution, including acoustic-emission monitoring[5,6,16–18], micropillar deformation[4,19], and nanoindentation[4,20,21], mean that it is now possible to probe the collective dynamics of populations of dislocations during plasticity and creep. In some cases, the flow behaviour appears smoothly continuous even at small length scales, termed mild plasticity[5,6,16]. Often however, deformation occurs in discrete bursts, implying the spatiotemporally correlated motion of groups of dislocations in avalanches, termed wild plasticity[5,6,16].

Whether a material exhibits mild or wild plasticity depends systematically on two key variables, specifically the length scale of observation and the resistance of the lattice to dislocation glide[6,16]. Over large length scales, the discrete motion of dislocations is difficult to detect and an averaged mild plastic behaviour is observed. At smaller length scales, individual bursts of dislocation motion are easier to detect and increasingly dominate the observed behaviour. Likewise, when lattice resistance is high, elastic interactions due to the motion of one dislocation have little impact on neighbouring dislocations that are effectively held in place. However, when lattice resistance is low, the motion of one dislocation can trigger an avalanche of dislocations in its vicinity.

These effects have important implications for the nature of plasticity and creep in the upper mantle, where both the extrapolation of laboratory flow laws[22] and observations of seismic anisotropy[23] indicate that deformation occurs predominantly by dislocation-mediated mechanisms. Whereas deformation of the mantle appears mild when observed by geodetic techniques[7–9] over length scales of $10^3$–$10^6$ m, plasticity and creep at smaller scales may be wilder. Furthermore, although classical experiments do probe smaller length scales on the order of $10^{-3}$–$10^{-2}$ m, at which plasticity again appears mild[10–12], the difference in strain rates between deformation in the laboratory and in nature generates large differences in effective lattice resistance between the two settings. A recently calibrated flow law for dislocation glide at high temperatures predicts that effective lattice resistance under the high temperatures and low strain rates of the asthenospheric upper mantle will be dramatically less than in laboratory experiments[24], increasing the potential for wild plasticity. Lastly, the pronounced viscoplastic anisotropy of olivine, which is the volumetrically dominant mineral in the upper mantle, limits the number of active slip systems, thereby reducing the potential for short-range dislocation interactions[25] that would increase the effective lattice resistance and promote mild plasticity[6,16]. Instead, long-range elastic interactions among dislocations dominate during plasticity[12,26] and creep[24,27,28] of olivine, again increasing the potential for wild plasticity[6,16].

These considerations suggest that the fundamental nature of plasticity and creep in the upper mantle may be different to intuition gained thus far from observations at large length scales or laboratory strain rates. Instead, the deformation of olivine should be evaluated within the framework of mild-to-wild plasticity to assess the underlying nature of collective dislocation dynamics in various laboratory and natural settings. This reappraisal will provide new constraints for rheological models based on dislocation motion and will provide a first step towards assessing the extent to which macroscopic instabilities during ductile flow, such as deep earthquakes and slow-slip events, may be linked to intermittent behaviours at smaller length scales.

Here, we analyse the deformation of olivine at the smallest length scales available in the laboratory using instrumented nanoindentation[29] to test whether intermittent plasticity occurs. These experiments reveal the occurrence of dislocation avalanches and allow olivine to be placed on the spectrum of mild-to-wild plasticity. This result allows the framework of mild-to-wild plasticity to make predictions of changes in the degree of wildness as the lattice friction is extrapolated to natural conditions. This analysis suggests that plasticity of the lithosphere is mild across most length scales. However, creep of the asthenosphere is likely wild, at least up to the grain scale and potentially beyond. These predictions constitute a previously unrecognised mild-to-wild transition with depth in Earth's upper mantle.

## Dislocation avalanches in nanoindentation experiments

Nanoindentation tests using a spheroconical indenter tip on single crystals of olivine with low defect densities exhibit three characteristic portions of the load-displacement curve and corresponding pseudo-stress-strain curve[29] as indicated in Fig. 1a. Initial loading is elastic and follows Hertzian contact mechanics. Elastic loading ends in a displacement burst, termed a 'pop-in', that marks the onset of plastic flow. Subsequent deformation proceeds by plastic flow and exhibits strain hardening. Both dislocations and microcracks are present in the deformed zone after each test[26]. However, both theoretical considerations and observational evidence indicate that, when indented by a tip with a radius on the order of a few micrometers or less, only dislocation motion occurs during loading, whereas microcracks initiate during unloading, even in brittle ceramics[29,30]. In olivine, the initial pop-in has been studied in detail to analyse dislocation nucleation and the onset of dislocation motion[29]. Instead, here we focus on the details of the third portion of the test to investigate dislocation dynamics during ongoing plastic flow.

Our key observation is that, after the initial pop-ins, the load-displacement curves exhibit smooth, continuous plastic flow that is punctuated by intermittent secondary displacement bursts. These bursts are evident as peaks in displacement rate in a typical test in Fig. 1a. Likewise, Fig. 1b presents the plastic-flow portion of the same test with the displacement bursts identified by our detection routine marked by black arrows. Segments of continuous flow proceed over displacements on the order of tens to hundreds of nanometres. In contrast, the magnitudes of displacement bursts have a mean of 5.4 nm, standard deviation of 2.9 nm, and maximum of 25.6 nm, but typically occur within one sampling time interval of 0.2 s. Displacement bursts were detected in 83% of tests and between 40 and 103 displacement bursts were detected in total for each sample.

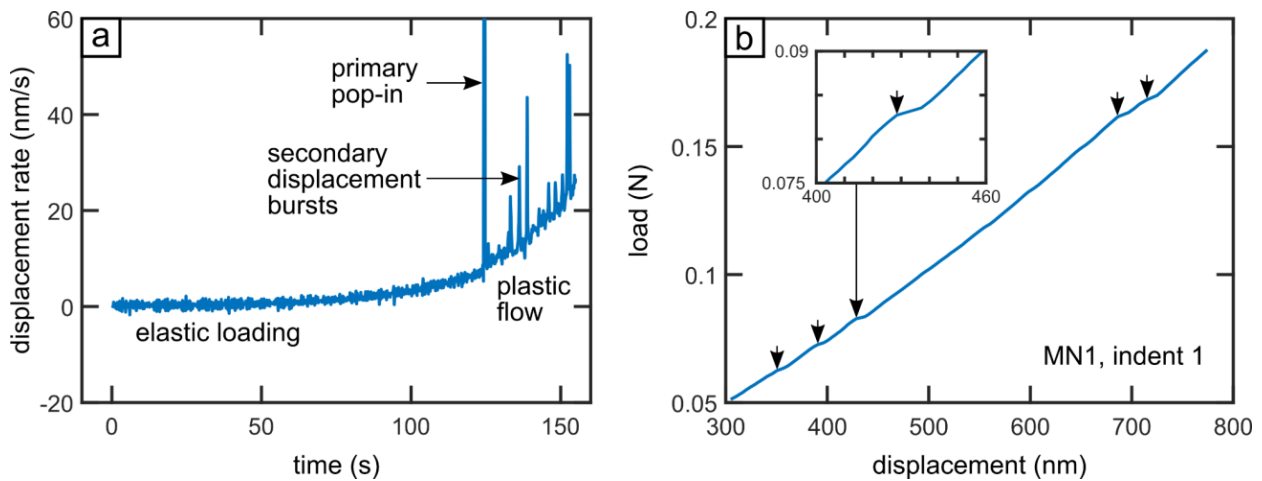

**Fig. 1 | Displacement bursts. a**, Typical time series of displacement rate indicating elastic loading and plastic flow separated by the primary pop-in, along with secondary displacement bursts during plastic flow. **b**, Typical load-displacement curve from the plastic-flow portion of the test after the primary pop-in. Black arrows indicate secondary displacement bursts that are detected and included in analysis. The inset displays an enlargement of the third displacement burst. Sample MN1, indent 1.

Analysis of the probability distributions of the magnitudes of displacement bursts provides additional insight into their characteristics. Fig. 2 presents normal probability plots, in which the cumulative-probability axis is scaled such that a normal distribution falls on a straight line. We plot the logarithm of the magnitudes of displacement occurring during bursts so that a straight line indicates a log-normal distribution. Fig. 2a reveals that the distributions from all samples are close to straight lines. This overall behaviour is clear in Fig. 2b, in which the 695 displacement bursts from across all samples are combined. The magnitudes of displacement bursts are log-normally distributed across the approximate range of 2–20 nm.

Similar log-normal distributions of secondary displacement bursts have recently been observed during spherical nanoindentation of Cu[20]. Cu does not fracture during nanoindentation with either spheroconical[31,32] or sharp[33,34] tips, providing further confirmation that secondary displacement bursts with these characteristics are formed by dislocation motion alone. In spheroconical nanoindentation, scale-invariant power-law distributions of displacement bursts, typical of correlated dislocation motion in many test systems[4–6,16,21,35], do not emerge because the size of the deforming volume and the dislocation density are evolving continuously throughout the test[20]. However, the absence of this behaviour does not imply that dislocation motion was mild, in which case Gaussian statistics would emerge[20]. Instead, the log-normal distribution of displacement-burst magnitudes is a signature of correlated dislocation motion in spheroconical nanoindentation[20], with displacement bursts of 2–20 nm in olivine resulting from avalanches of up to approximately several tens of dislocations.

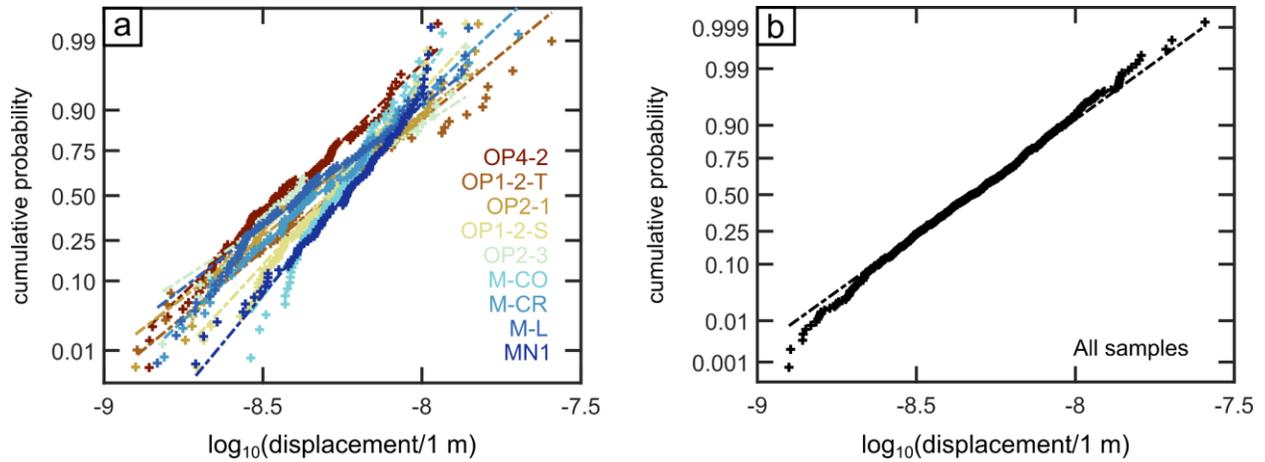

**Fig. 2 | Probability distributions of displacement bursts. a**, Normal probability plot of the logarithm of magnitudes of secondary displacement bursts in each sample. The vertical axis is scaled such that a normal distribution plots as a straight line. As we take the logarithm of the displacement, datasets that plot as a straight line are log-normally distributed. Dashed lines indicate straight lines fitted to the distributions from each sample. **b**, Normal probability plot of the logarithm of magnitudes of secondary displacement bursts in all samples.

A key advantage of spherical nanoindentation is the capability to determine pseudo-stress-strain curves from load-displacement curves[36] (Methods), allowing estimates of plastic strain accommodated during secondary displacement bursts. Across all samples, the mean plastic strain accommodated during each displacement burst is estimated to be $1.3 \times 10^{-3}$, with a standard deviation of $1.1 \times 10^{-3}$ and a maximum of $1.0 \times 10^{-2}$. The fraction of the total plastic strain after the initial pop-in that is accommodated in secondary displacement bursts provides a quantitative estimate of wildness. This fraction varies between 4% and 12% among samples, with an average wildness of $8 \pm 6\%$ (one standard deviation) across all tests.

The framework of mild-to-wild plasticity allows comparisons among the wildness of different materials. In general, wildness varies as a function of a ratio, $R$, of length scales, $R = L/l$, where $L$ is the system size and $l$ is an internal length scale[6,16]. This internal length scale quantifies the relative resistance to dislocation motion from long-range dislocation interactions and short-range obstacles through $l = Gb/\tau$, where $G$ is the shear modulus, $b$ is the magnitude of the Burgers vector, and $\tau$ is the combined resistance imposed by the lattice, short-range dislocation interactions, solution hardening, and/or precipitates[6,16]. In general, when $R$ is large (i.e., when the system is large or strengthening from short-range obstacles dominates) plasticity appears mild, whereas when $R$ is small (i.e., when the system is small or long-range interactions dominate) plasticity appears wild[6,16]. Fig. 3 presents this trend for several different materials.

Plasticity of olivine can also be considered in this context. We estimate the diameter of the deforming region beneath the indenter tip to be approximately 5 μm based on transmission electron micrographs[26,27]. Detailed experimental[12,24] and microstructural[26,27,37] analyses suggest that strain hardening in olivine is dominated by long-range elastic interactions with negligible contributions from short-range interactions, solution strengthening, or precipitates. Therefore, we take $\tau$ to be a lattice resistance at room temperature of 1.4 GPa[12], giving $R = 167$. This high value of $R$ and low values of wildness for olivine at room temperature are broadly consistent with the trend defined by other materials in Figure 3.

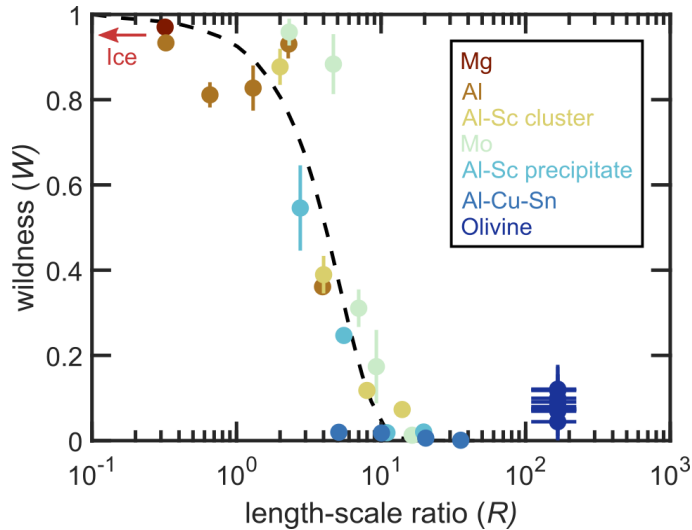

**Fig. 3 | Wildness.** Wildness against the dimensionless length-scale ratio. Wildness is the fraction of plastic strain occurring in secondary displacement bursts. The length-scale ratio is computed using a shear modulus of 70 GPa, Burgers vector of 0.6 nm, and lattice resistance of 1.4 GPa[12]. Modified from Weiss et al.[5,6].

# Mild-to-wild plasticity in the upper mantle

Having demonstrated that olivine displays dislocation avalanches at room temperature, and that those avalanches can be considered in the wider framework of mild-to-wild plasticity, we can use that framework to predict the behaviour of olivine at higher temperatures and slower strain rates[12,24]. The principal effect of increasing temperature and decreasing strain rate will be to reduce the lattice resistance, which in turn will increase the wildness of plasticity. Fig. 4 presents predictions of the length-scale ratio, $R$, as a function of lattice resistance, $\tau$, and system size, $L$, with the fields dominated by mild and wild plasticity approximately demarcated by the $R = 1$ line. Fig. 4 also indicates the approximate loci of various experiments and regions of the upper mantle. These predictions are made possible by recent measurements[24] of lattice resistance to glide at temperatures up to 1300°C. Plasticity of the lithospheric upper mantle at temperatures of approximately 800°C is predicted to appear mild at all but the shortest length scales due to high lattice resistance at low temperatures. However, plasticity in the asthenospheric upper mantle at temperatures of around 1300°C is predicted to be extremely wild due to low lattice friction at high temperature and slow strain rates. Notably, plasticity should appear wild at length scales at least up to the grain size.

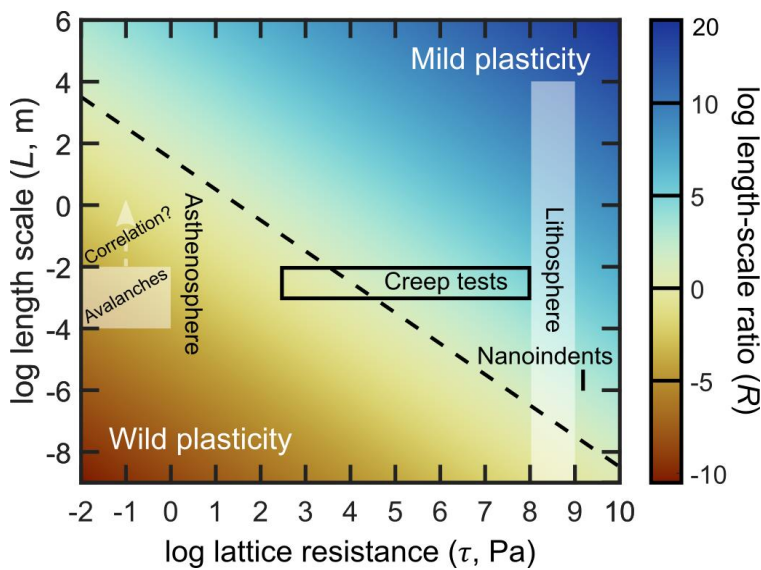

**Fig. 4 | Predicted regimes of mild and wild plasticity.** The dimensionless ratio ($R$) between the length scale of observation ($L$) and characteristic microstructural length scale ($l$) as a function of the length scale of observation and the lattice resistance to dislocation glide ($\tau$). The length scale and lattice resistance of the nanoindentation experiments were estimated from Wallis et al.[26] and Hansen et al.[12], respectively. The approximate range of lattice resistances during high-temperature creep tests was estimated from Hansen et al.[24] based on a strain rate of $10^{-5}$ s$^{-1}$, dislocation density of $10^{12}$ m$^{-2}$, and temperatures in the range 1200–1600°C. The approximate ranges of lattice resistances occurring in the lithosphere and asthenosphere were estimated from Hansen et al.[24] based on a strain rate of $10^{-14}$ s$^{-1}$, dislocation density of $10^{10}$ m$^{-2}$, and temperatures in the range 800–1400°C. The box labelled 'avalanches' indicates the approximate lattice resistance and grain-size range estimated for the asthenospheric mantle. This grain-size range likely limits the maximum size of individual dislocation avalanches but avalanches may exhibit spatiotemporal correlations extending to greater length scales indicated approximately by the arrow labelled 'correlation'. The dashed black line marks $R = 1$, separating regimes of dominantly mild and wild plasticity above and below, respectively.

At the high homologous temperatures and slow strain rates of the asthenosphere, the behaviour of olivine may be analogous to that of water ice, which is the wild endmember of plasticity due to its low lattice friction and general lack of short-range dislocation interactions[16] (Fig. 3). Acoustic-emission monitoring of dislocation avalanches in polycrystalline ice has revealed that individual dislocation avalanches are confined within individual grains[16,17]. However, avalanches exhibit spatial and temporal correlations over distances greater than the grain size, indicating that avalanches can trigger cascades in nearby grains[16–18]. It remains an open question as to how large these correlated dislocation avalanches may become in the asthenosphere, where the system size is large and long timescales are available to capture low-probability events.

Overall, these considerations suggest that the upper mantle likely exhibits a mild-to-wild transition with increasing depth and temperature from the lithosphere into the asthenosphere. The wild plasticity predicted at high temperatures is a striking departure from the current paradigm of slow continuous creep. The predictions in Fig. 4 suggest that the existence of this mild-to-wild transition in olivine is testable in laboratory creep experiments, which should span the transition if conducted over temperatures in the range 1200–1600°C. In such tests, dislocation avalanches may be detectable as acoustic emissions, which are predicted to become more intermittent and pronounced with increasing temperature. We also note that nanoindentation creep tests on halite[38] exhibit transient periods of dramatically accelerated deformation that provide an example of instabilities during power-law dislocation creep at homologous temperatures higher than those of our experiments.

Our findings raise intriguing questions about the nature of plasticity and viscous flow of olivine in the upper mantle and of other minerals in other geological contexts. Considering that our nanoindentation tests indicate that dislocation avalanches occur in olivine at room temperature (Figs. 1 and 2), that the framework of mild-to-wild plasticity predicts that deformation should get wilder with depth in Earth (Fig. 4), and that dislocation avalanches generally exhibit spatiotemporal correlations that effectively extend the range of the instability[16–18], it is plausible that that transient phenomena associated with the viscous regime, such as deep earthquakes[13,14] and slow-slip events[15], may be triggered by or wholly result from these or related effects.

## Methods

### *Nanoindentation data acquisition and processing*

The spherical nanoindentation tests in this study were originally reported by Kumamoto et al.[29], and further details and data are available there. The methods used for acquiring and processing the data are also described here. Tests were performed using an MTS XP Nanoindenter and spheroconical diamond indenter tip. In all tests, nominal strain rate (loading rate divided by load) was kept constant at $0.05$ s$^{-1}$.

Nanoindentation tests were first performed on a fused silica standard to determine the effective radius of the tip. We performed 16 purely elastic indents with maximum displacements into the sample in the range 40–70 nm. The effective modulus, $E_{\text{eff}}$, of the indentation test, which takes into account the elasticity of both the sample and the indenter tip, was calculated as

$$\frac{1}{E_{\text{eff}}} = \frac{1-\nu_s^2}{E_s} + \frac{1-\nu_i^2}{E_i},$$

Eq. S1

where $E$ is Young's modulus, $\nu$ is Poisson's ratio, and the subscripts i and s refer to the indenter tip and the sample, respectively. The effective radius of the indenter tip, $R_{\text{eff}}$, was then calculated using the equation for a Hertzian contact as

$$P = \frac{4}{3} E_{\text{eff}} R_{\text{eff}}^{1/2} h^{3/2}.$$

Eq. S2

In this equation, $P$ is load and $h$ is displacement of the tip into the sample surface. As the indents in fused silica were purely elastic, $R_{\text{eff}}$ is taken to be equal to the true indenter radius.

Nanoindentation experiments on olivine were carried out using a tip with an effective radius of 3 μm. After using the method of Kalidindi and Pathak[39] to determine the effective touch point for each test, $E_{\text{eff}}$ was calculated for each crystal by fitting the Hertzian contact relationship to purely elastic indents (70 nm deep) using the calibrated indenter radius.

We use Kalidindi and Pathak's[39] definitions of contact radius, $a$, and strain, $\varepsilon$, as they produce curves that are good analogs of stress-strain curves. As our experiments were performed using continuous stiffness measurement, the resulting contact radius, hardness, $H$, and strain could be calculated at all points in the test as

$$a = \frac{S}{2E_{\text{eff}}},$$

Eq. S3

$$H = \frac{P}{\pi a^2},$$

Eq. S4

and

$$\varepsilon = \frac{4}{3\pi}\frac{h}{a}. \hspace{5em} \text{Eq. S5}$$

The precision of load and displacement measurements on the MTS XP Nanoindenter are 50 nN and < 0.01 nm, respectively, according to manufacturer specifications. However, the precision of a real measurement is driven as much by the laboratory environment as the machine specifications. A reasonable approximation of the precision can be estimated using a running standard deviation and running average. Such estimates should be conservative as all values of interest (load, displacement, stiffness, hardness, and strain) generally increase throughout the experiment. The precision of the three machine-generated values (load, displacement, and stiffness) is ~1–2%. The precision of hardness and strain are ~2–3% and 1–2%, respectively.

Surface-breaking fractures are visible around most indents after the experiments[26]. However, both theoretical considerations and observational evidence indicate that, when indented by a tip with a radius on the order of a few micrometers or less, only dislocation motion occurs during loading, whereas microcracks initiate during unloading, even in brittle ceramics[29,30,40]. Likewise, examination of the fracture structure using focused ion beam milling suggested these fractures result from the unloading process as residual stresses are released[29]. These interpretations are supported by the similar log-normal forms of the distributions of secondary displacement bursts occurring during spherical nanoindentation of olivine (Fig. 2) and Cu[20]. As Cu does not fracture during nanoindentation with either spheroconical[31,32] or sharp[33,34] tips, this similarity provides further confirmation that secondary displacement bursts with these characteristics are formed by dislocation motion alone.

## *Analysis of displacement bursts*

We analysed displacement bursts in the plastic-flow portion of each test after the larger initial displacement burst of the primary pop-in. To automatically detect displacement bursts in the load-displacement curve, we detected peaks in displacement/load as a function of displacement. To set the threshold for peak detection in the plastic-flow portion of each test, we analysed changes in the gradient of the load-displacement curve during the elastic-loading portion of the same test. We assume that fluctuations in displacement/load during elastic loading arise from instrument noise, temperature variation, and/or other factors unrelated to plasticity. We measured the prominences of peaks in the displacement/load series during the final 10 s of elastic loading before the primary pop-in and took three standard deviations of these peak prominences to be the threshold prominence for peak detection in the displacement/load series during the plastic-flow portion of the same test. That is, displacement bursts were detected in the plastic-flow portion of a test if they were more pronounced than 99% of the fluctuations occurring during the latter stages of elastic loading. The detected displacement bursts were visually checked on each load-displacement curve (e.g., Fig. 1).

To estimate the plastic strain that occurred during each displacement burst, we analysed the indentation hardness-strain series during the plastic-flow portion of each test. To estimate differential stress during indentation experiments, the hardness is often down scaled by a factor termed the constraint factor to account for the confining effect of the sample surrounding the deforming zone[41]. The value of this constraint factor ranges from 1 for a perfectly elastic material to 3 for a perfectly plastic material[41]. However, the determination of the appropriate constraint factor to use in spherical nanoindentation is complex and poorly established[42,43], especially for materials, such as olivine, that exhibit elastic anisotropy[44], plastic anisotropy[12,29], and kinematic hardening by the generation of back stress[12]. Therefore, we take a simple and conservative approach of assuming a constraint factor of 1 and thereby correct for elastic effects based on changes in hardness. In doing so, we maximise elastic-strain corrections to the total strain, resulting in conservative, lower-bound estimates of average plastic strain during displacement bursts and the wildness of each sample. To perform this correction for elastic effects, we used the Young's modulus specific to each sample[29]. Furthermore, if the true duration of a displacement burst was less than the 0.2 s sampling interval then the plastic strain during the displacement burst will be overestimated due to a contribution of steady plastic flow also occurring during the sampling interval. For a typical rate of steady plastic flow on the order of $1.5 \times 10^{-3}$ s$^{-1}$, a plastic strain of $3 \times 10^{-4}$ would accrue in 0.2 s by steady plastic flow if the displacement burst was instantaneous. As the mean plastic strain occurring during the 0.2 s sampling intervals associated with the displacement bursts is $1.3 \times 10^{-3}$, these strains may overestimate those actually occurring during the displacement bursts by up to approximately 20% (i.e., the strain contributed instead by steady plastic flow) depending on the true durations of the displacement bursts.

## Acknowledgements

This work was supported by the Natural Environment Research Council [grant number NE/M000966/1]; a UK Research and Innovation Future Leaders Fellowship [grant number MR/V021788/1]; and the Netherlands Organisation for Scientific Research, User Support Programme Space Research [grant numbers ALWGO.2018.038 and ENW.GO.001.005]. Portions of this work were performed under the auspices of the U.S. Department of Energy by Lawrence Livermore National Laboratory under Contract DE-AC52-07NA27344. LLNL-JRNL-852165

## Author contributions

DW conceived the study and analysed the data. KMK collected the data. Both authors wrote the manuscript.